\documentstyle{mn}
\input{epsf.sty}
\newcommand{\figpath}{.}
\def\plottwo#1#2{\centering \leavevmode
\epsfxsize=.95\columnwidth \epsfbox{#1} \hfil
\epsfxsize=.95\columnwidth \epsfbox{#2}}
\def\plotsmall#1{\centering \leavevmode \epsfxsize=0.95\columnwidth \epsfbox{#1}}
\begin{document}

\title[Non-coplanar disc-disc encounters]{Numerical simulations of protostellar encounters \\ III. Non-coplanar disc-disc encounters}

\author[S.J.~Watkins et al.]
  {S.J.~Watkins,
  A.S.~Bhattal,
  H.M.J.~Boffin,
  N.~Francis
  and A.P.~Whitworth\\
  Department of Physics and Astronomy, University of Wales, Cardiff CF2 3YB, Wales, UK.
  }

\maketitle
\begin{abstract}

It is expected that an average protostar will undergo at least one impulsive interaction with a neighbouring protostar whilst a large fraction of its mass is still in a massive, extended disc. If protostars are formed individually within a cluster before falling together and interacting, there should be no preferred orientation for such interactions. As star formation within clusters is believed to be coeval, it is probable that during interactions, {\it both} protostars possess massive, extended discs.

We have used an SPH code to carry out a series of simulations of non-colpanar disc-disc interactions. We find that non-coplanar interactions trigger gravitational instabilities in the discs, which may then fragment to form new companions to the existing stars. (This is different from coplanar interactions, in which most of the new companion stars form after material in the discs has been swept up into a shock layer, and this then fragments.) The original stars may also capture each other, leading to the formation of a small-${\cal N}$ cluster. If every star undergoes a randomly oriented disc-disc interaction, then the outcome will be the birth of many new stars. Approximately two-thirds of the stars will end up in multiple systems.

\end{abstract}
\begin{keywords}
stars: formation -- binaries: general -- accretion, accretion discs -- methods: numerical -- hydrodynamics -- instabilities
\end{keywords}

\section{Introduction}

Interactions between young protostars are believed to play an important r\^{o}le in the star formation process, with particular implications for the formation of binary systems, and for the lifetimes of discs. Larson \shortcite{larson90} suggested that capture during interactions might be an important binary formation mechanism, but further investigations (Clarke \& Pringle 1991a, Clarke \& Pringle 1991b, Hall, Clarke \& Pringle 1995, Heller 1995) indicated that capture was unlikely to be a dominant mechanism.
 
In all of the above cases, only interactions in which just one of the stars possessed a disc were investigated. However, observations indicate that stars within clusters tend to be formed coevally (e.g.\ Gauvin \& Strom 1992), and so it might be expected that both interacting stars would possess extended discs. In particular, if star formation is triggered dynamically, then the stars formed are likely to have a small spread of ages, and so again disc-disc interactions should tend to predominate over star-disc interactions. 

Numerical simulations of star formation triggered by the collision of two clumps within a molecular cloud (Turner et al.\ 1995\nocite{turner}, Whitworth et al.\ 1995\nocite{whitworth}) suggest that the majority of protostars will undergo interactions whilst still surrounded by massive, extended discs, and that most of these interactions may be prograde and coplanar. If, however, star formation is more distributed, and interactions occur when two protostars formed separately within a cluster fall together, then there should be no preferred orientation for interactions.

This is the last of three papers describing the results of simulations of protostellar interactions. Paper I \cite{watkins97a} described interactions in which only one of the stars possessed a disc, and Paper II \cite{watkins97b} presented the results of simulations of coplanar interactions in which both protostars had massive, extended discs. This paper investigates non-coplanar interactions between two stars each possessing a massive, extended disc, analyses the results presented in both this paper and Paper II, and compares them with the results for star-disc interactions presented in Paper I.

\section{Physical and computational model}

The numerical method used is described in Paper I. The discs modelled in the simulations described in this paper have the same physical properties as the discs used in Paper I. A brief summary of these properties is given in Table \ref{table:phys}. Each disc is modelled with 2000 SPH particles, so it is not resolved vertically. One particle represents a mass of $2.5 \times 10^{-4} M_\odot$.

\begin{table}
\begin{tabular}{l|l}
property & value   \\ \hline
disc radius & 1000AU \\
disc mass & 0.5$\mbox{M}_{\sun}$ \\
star mass & 0.5$\mbox{M}_{\sun}$ \\
disc density profile & $r^{-3/2}$ \\
eccentricity of orbit & 1.0 \\
initial separation & 5000AU \\
shear viscosity & $\nu_{s} = c_{s} h /800$ \\
\end{tabular}
\caption[Physical parameters used in simulations]
{\label{table:phys}Physical parameters used in simulations}
\end{table}

 Non-coplanar disc-disc interactions are simulated at the same periastra and orbit inclinations as the star-disc encounters, in order to limit the parameter space to be searched. In all cases, the spin of the secondary disc is parallel to the orbit. A total of 12 simulations of non-coplanar disc-disc interactions has been carried out.

The simulations are labelled so that the number of the disc-disc simulation is the same as that of the corresponding disc-star simulation from Paper I, so that for instance the disc-disc simulation dd05 has the same orbital inclination and periastron as the disc-star simulation ds05.

\begin{table}
\caption[]{\label{table:runs}
List of simulations}
\begin{tabular}{|c|c|c|c|c|r|r|} \hline \hline
Run    & $r_{peri}$ & $\phi$   \\ \hline \hline
dd05-08 & 500, 1000, 1500, 2000 & $\phi=\pi/4$ \\ \hline
dd09-12 & 500, 1000, 1500, 2000 & $\phi=\pi/2$ \\ \hline
dd13-16 & 500, 1000, 1500, 2000 & $\phi=3\pi/4$ \\ \hline
\end{tabular}
\end{table}

\section{Simulations}

\begin{figure*}
\plottwo{\figpath/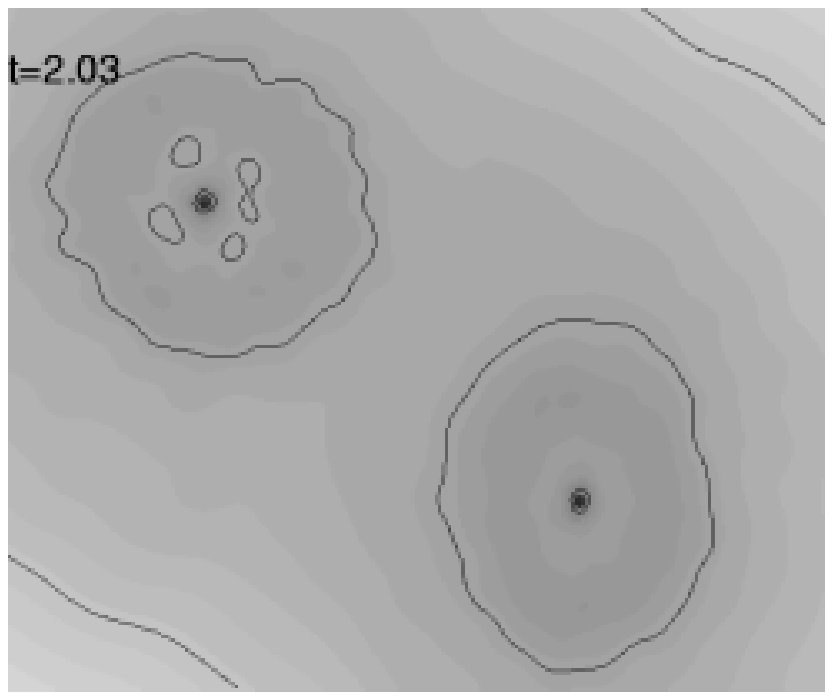}{\figpath/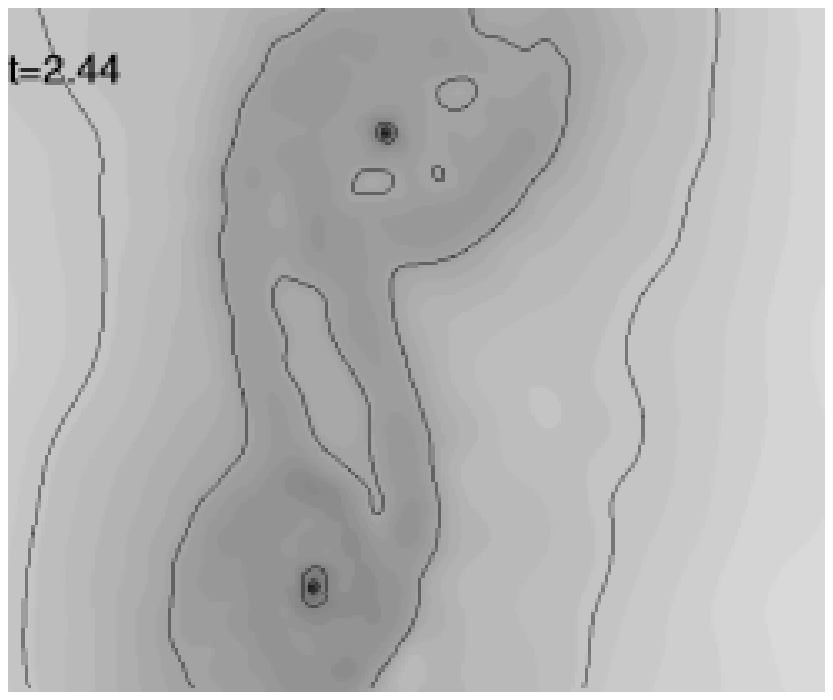}\\
\vspace{6pt}
\plottwo{\figpath/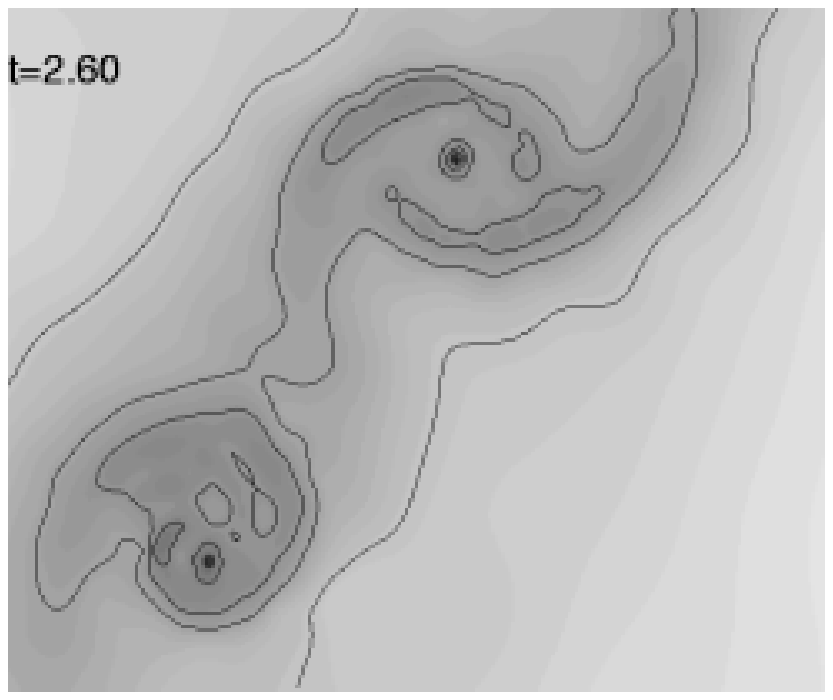}{\figpath/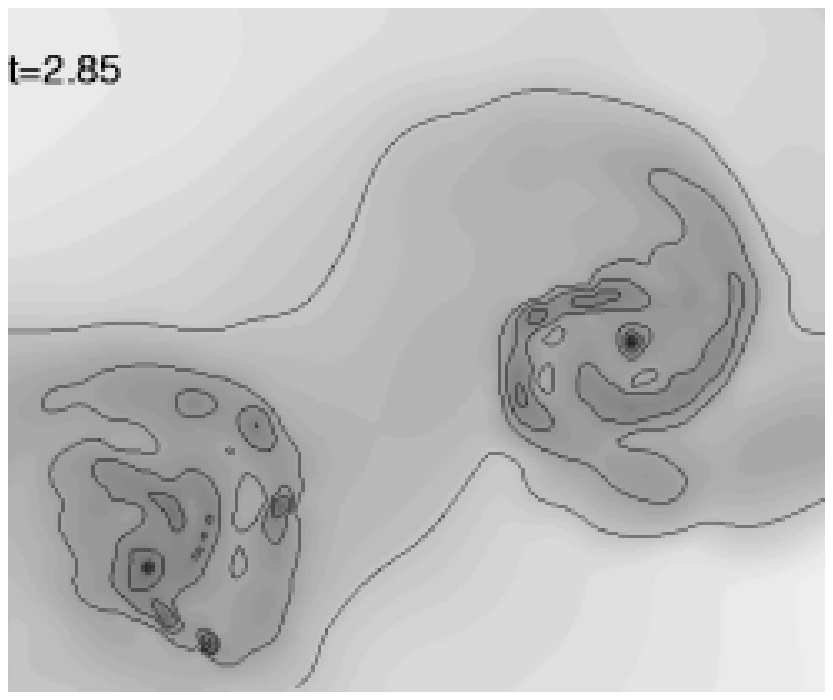}\\
\vspace{6pt}
\plottwo{\figpath/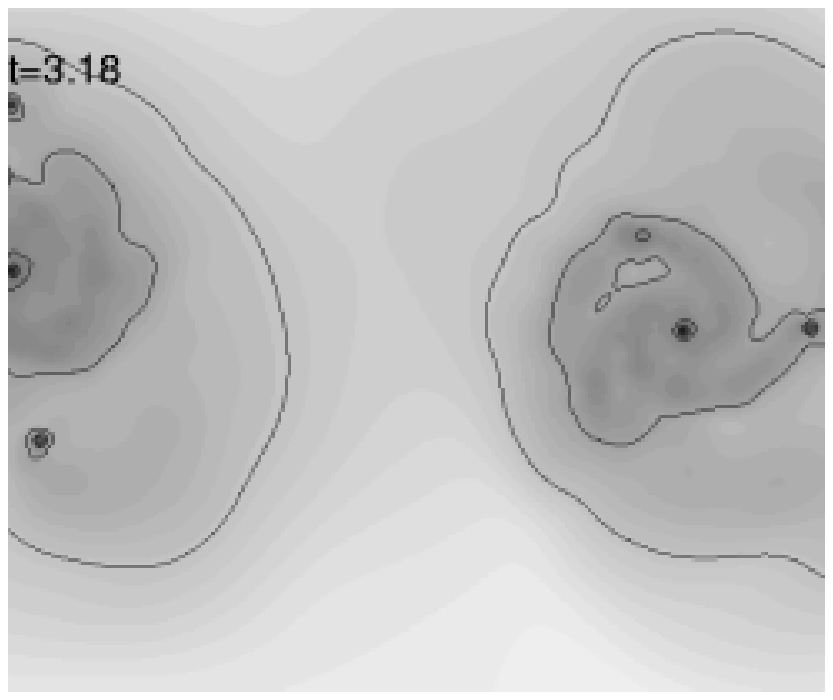}{\figpath/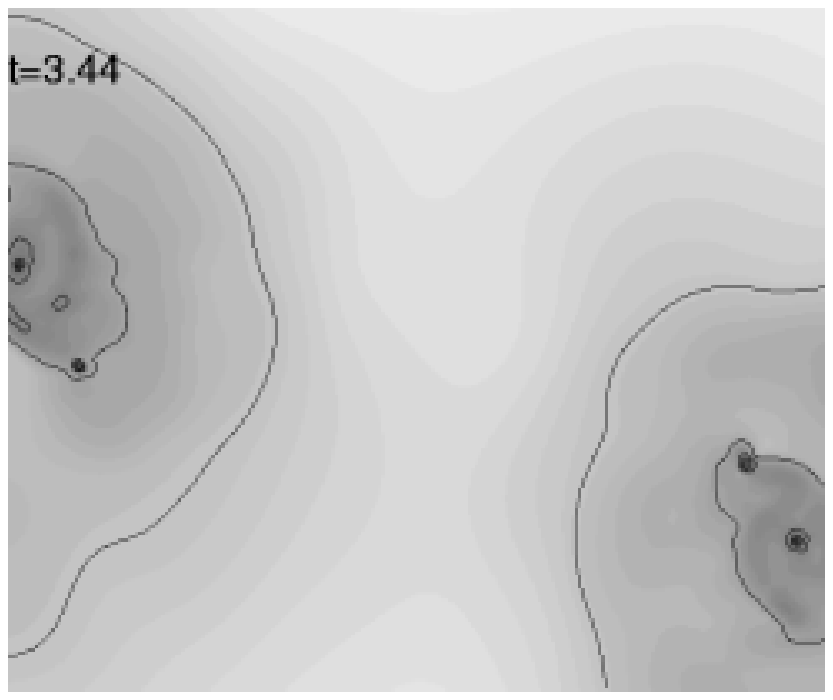}\\
\caption[Simulation dd06 : $\phi=\pi/4$, $r_{peri}=1000 \mbox{AU}$. 2000 $\times$ 1500 AU region, $x-y$ plane.]{\label{fig:dd06}Simulation dd06 : $\phi=\pi/4$, $r_{peri}=1000 \mbox{AU}$. 2000 $\times$ 1500 AU region, $x-y$ plane. Contour levels at [0.3, 3, 9 (frame 3, 4), 30, 300] $\mbox{g cm}^{-2}$. The primary is initially on the left of the figure, rotating clockwise and in the plane $z=0$. The secondary is initially below and to the right of the primary. Both the orbit and the secondary are inclined at $\pi/4$ to the primary. Viewed from this angle, the orbit is in a clockwise sense. As they interact, spiral arm instabilities cause the fragmentation of the discs, resulting in the formation of a binary and a triple system.}
\end{figure*}

The figures presented in this section show the results of three of the simulations conducted. The figures are grey-scale plots of surface density, in which the shading is logarithmically-scaled. These plots are overlaid with contours of constant surface density, that are equally separated in log-space. Occasionally an extra contour has been added in order to highlight an important feature of the figure, such as a shock or spiral arm. The contour levels used are given in the caption to the figure. The time at each figure, in units of $10^{4}$ years, is also given, and is calculated from the start of the simulation, when the two stars are at a separation of 5$r_{disc}$. The individual frames within a figure, when referred to within the text or a caption, are labelled from top left to bottom right in rows, so that frames 1 and 2 are on the top row, frames 3 and 4 on the next row and so on.

\subsection{run dd06 : $\phi=\pi/4$}

In this simulation, the orbit and the spin of the secondary are inclined at 45 degrees to the spin of the primary, and the periastron is 1000AU. The secondary starts below the plane of the primary, and passes up through it. There is no significant shock as the secondary disc penetrates through the primary. Instead the evolution of the system is dominated by a tidal-arm instability. A spiral arm forms joining the secondary disc to the primary star, and likewise joining the primary disc to the secondary star. Corresponding opposing tails form in the discs. The arms joining the stars are quickly dissipated and merge with the central discs surrounding each star, while the tidal tails remain in a circumbinary disc-like distribution. The inner regions of the discs undergo strong spiral-arm instabilities that lead to their fragmentation. The disc surrounding the primary produces one new condensation, whilst that around the secondary produces three. Two of these fragments merge, so that at the end of the simulation the primary is in a binary and the secondary is in an unstable triple system. The primary has a circumstellar disc of mass 0.18$\mbox{M}_{\sun}$ and radius 160AU, whilst its companion is of mass 0.06$\mbox{M}_{\sun}$. The binary orbit has eccentricity 0.14 and periastron 185AU. The secondary has a circumstellar disc of mass 0.24$\mbox{M}_{\sun}$ and radius 110AU. The companions are accreting from this disc and truncating it. The companions have masses 0.04$\mbox{M}_{\sun}$ and 0.09$\mbox{M}_{\sun}$. The binary and the triple system are bound to each other, with an eccentricity of 0.7 and a periastron of 860AU. The primary disc has been twisted through an angle of $9^{\circ}$, and its binary companion is orbiting in the plane of the disc. The secondary disc has been twisted though $5^{\circ}$, and likewise its companions lie in the plane of the disc.

\begin{figure*}
\plottwo{\figpath/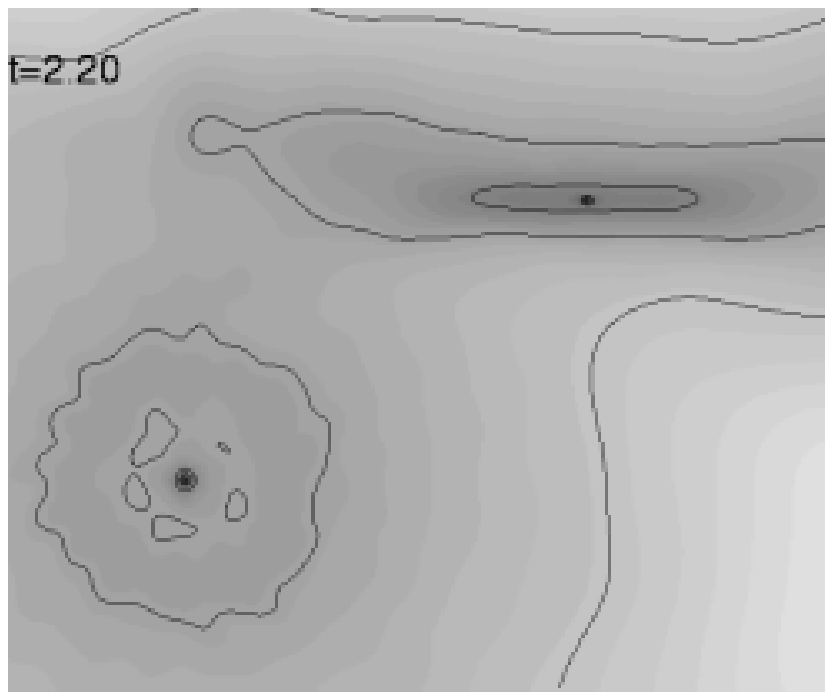}{\figpath/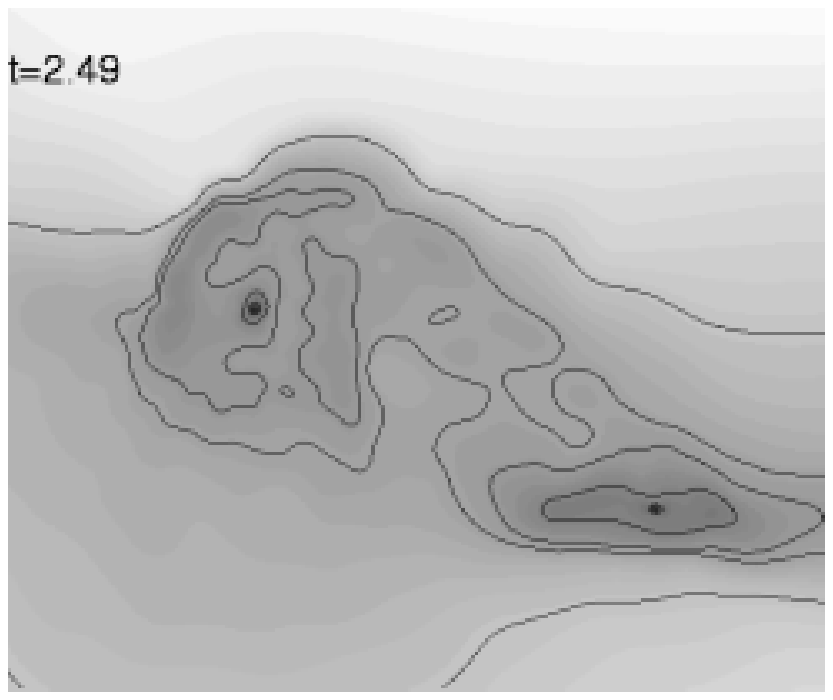}\\
\vspace{6pt}
\plottwo{\figpath/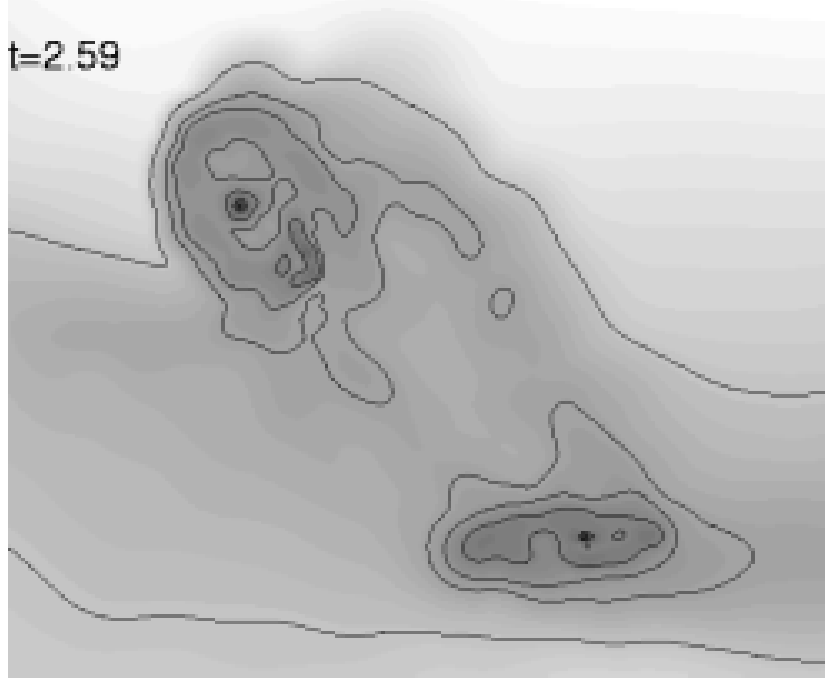}{\figpath/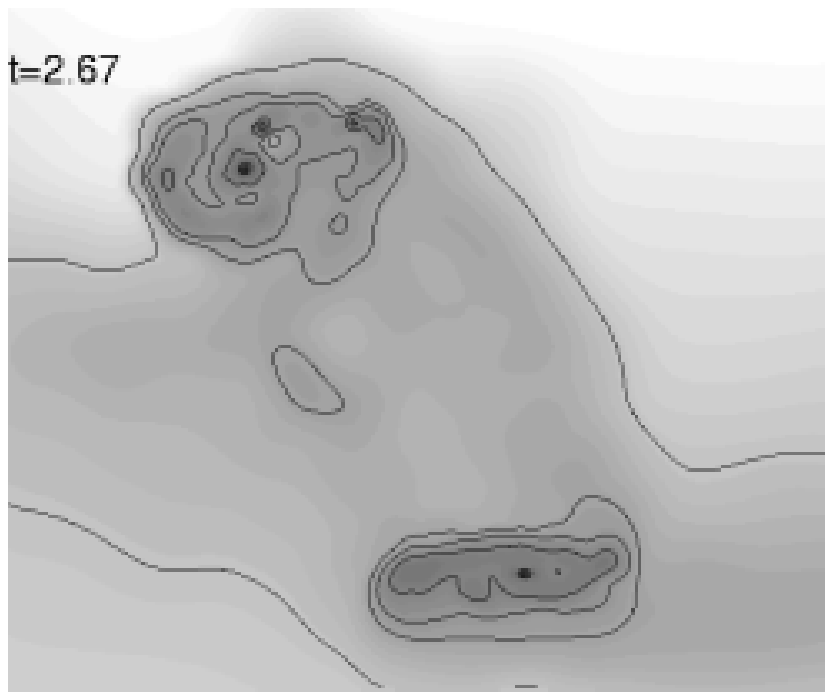}\\
\vspace{6pt}
\plottwo{\figpath/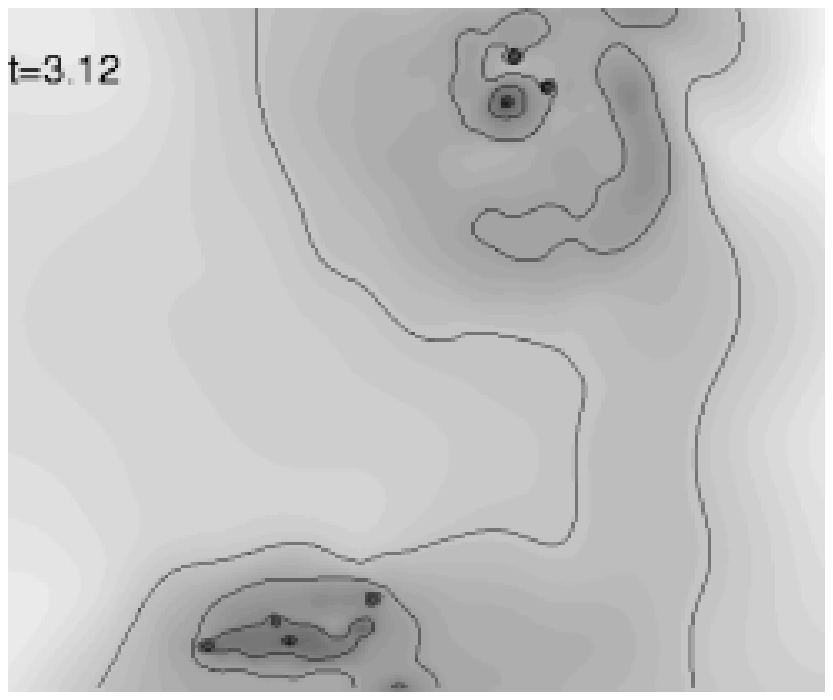}{\figpath/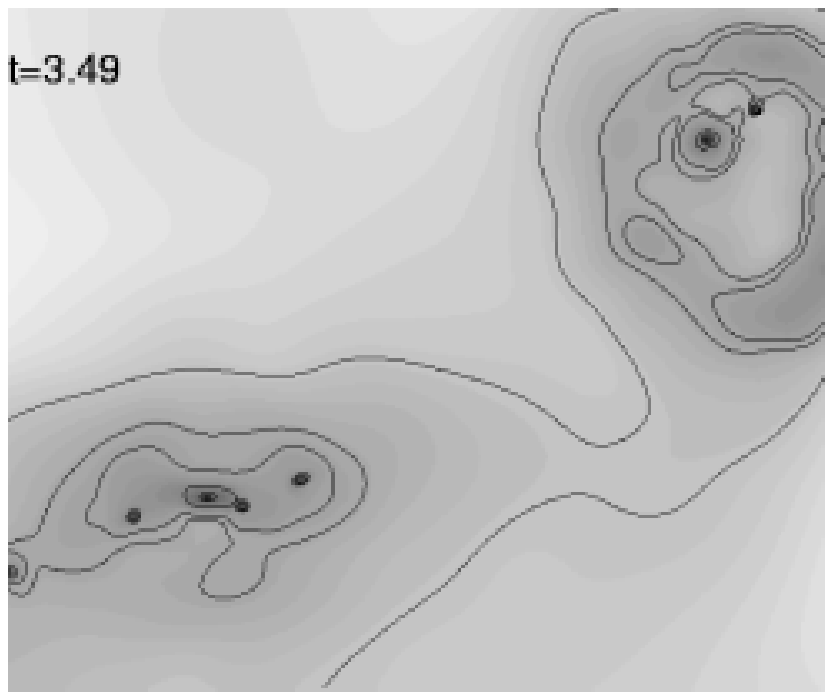}\\
\caption[Simulation dd10 : $\phi=\pi/2$, $r_{peri}=1000 \mbox{AU}$. 2000 $\times$ 1500 AU region, $y-z$ plane.]{\label{fig:dd10a}Simulation dd10 : $\phi=\pi/2$, $r_{peri}=1000 \mbox{AU}$. 2000 $\times$ 1500 AU region, $y-z$ plane. Contour levels at [0.3, 2 (frame 6), 3, 6 (frame 2), 7 (frame 3, 4), 30, 300] $\mbox{g cm}^{-2}$. The primary is initially at the top of the figure and is viewed edge-on. The secondary is initially below the primary. Viewed from this angle, the secondary is rotating clockwise, and the orbit of the encounter is also clockwise. The secondary disc fragments to form a binary, whilst the primary disc fragments to produce four new low-mass protostars.}
\end{figure*}

\begin{figure*}
\plottwo{\figpath/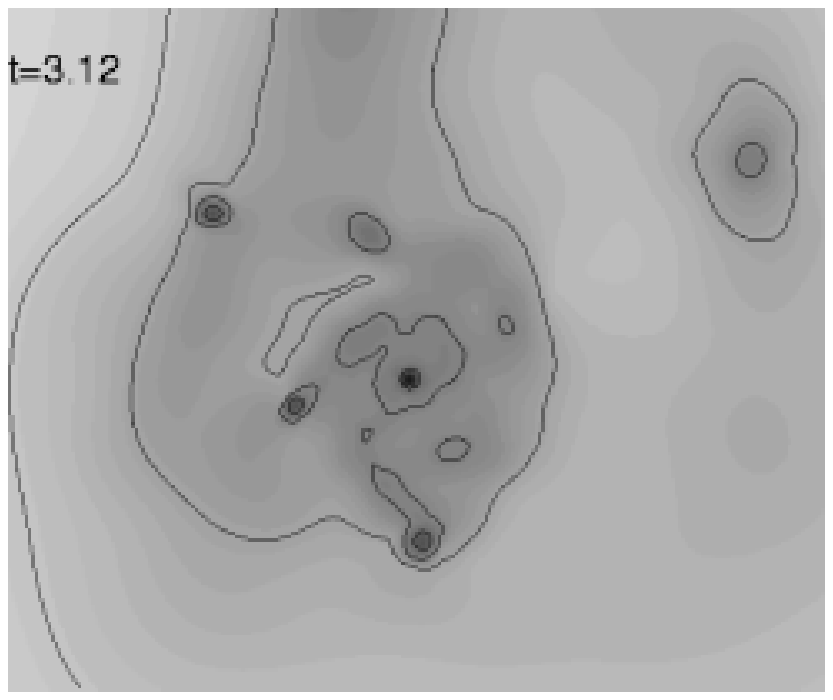}{\figpath/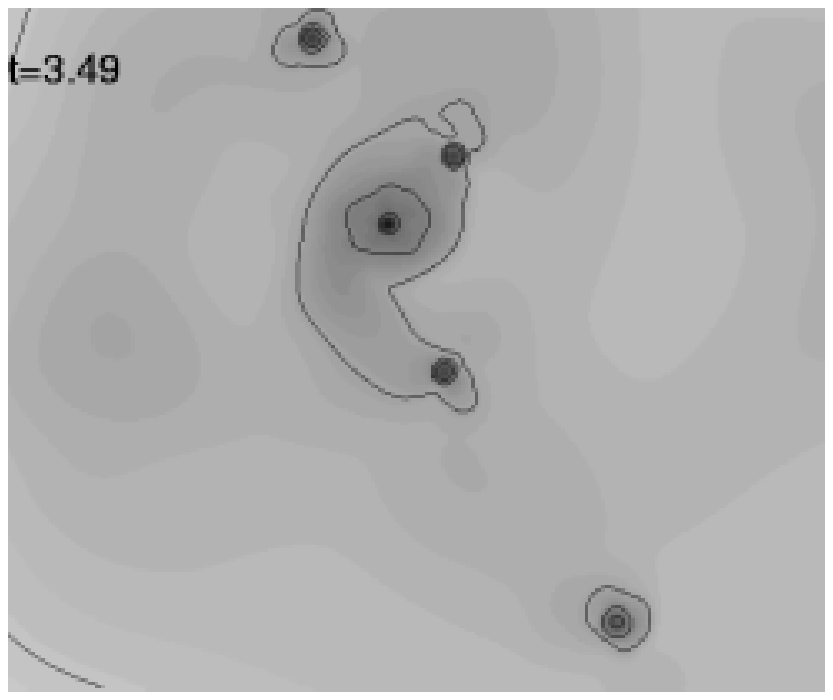}\\
\caption[Simulation dd10 : $\phi=\pi/2$, $r_{peri}=1000 \mbox{AU}$. 1000 $\times$ 750 AU region, $x-y$ plane.]{\label{fig:dd10b}Simulation dd10 : $\phi=\pi/2$, $r_{peri}=1000 \mbox{AU}$. 1000 $\times$ 750 AU region, $x-y$ plane. Contour levels at [0.3, 3, 30, 300] $\mbox{g cm}^{-2}$. View from above showing fragmentation of the primary disc to produce four new protostars.}
\end{figure*}

\begin{figure*}
\plottwo{\figpath/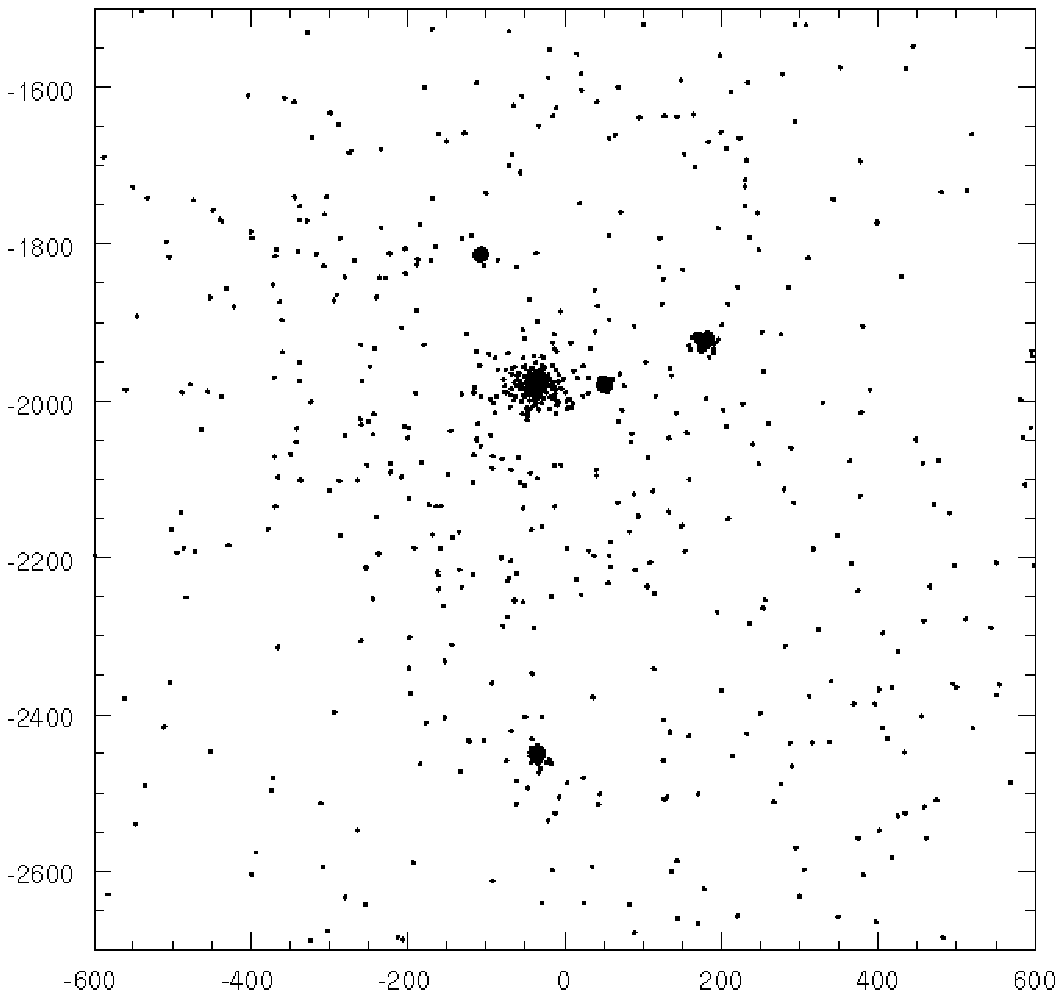}{\figpath/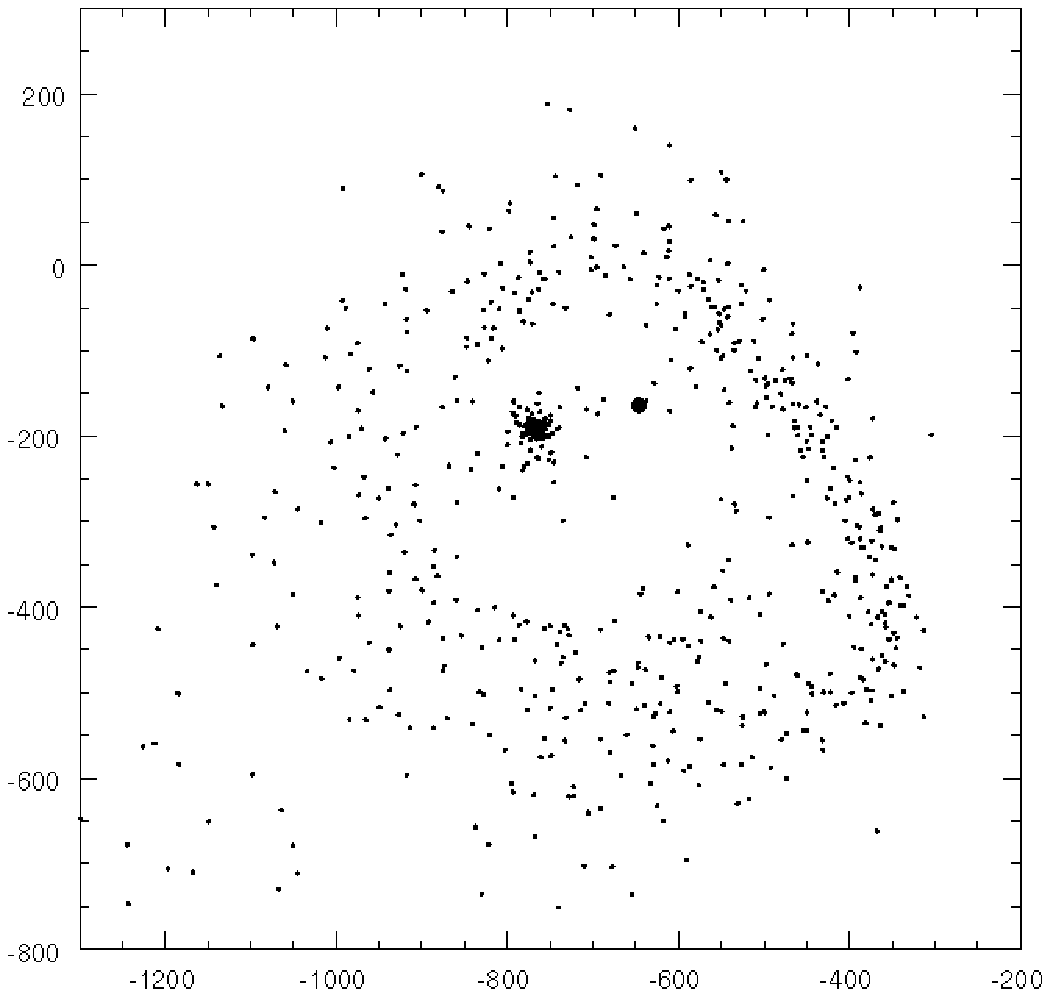}\\
\caption[Simulation dd10 - particle plot]{\label{fig:dd10c}Simulation dd10 : $\phi=\pi/2$, $r_{peri}=1000 \mbox{AU}$. Positions of SPH particles at end of simulation. The left-hand figure shows the primary star and its four companions. The right-hand figure shows the secondary star, its binary companion and the circumbinary disc in which they have cleared a gap.}
\end{figure*}

\subsection{run dd10 : $\phi=\pi/2$}

Figure \ref{fig:dd10a} shows the results for run dd10, in which the periastron is 1000AU. The secondary is initially below the primary, and orthogonal to it. As the secondary slices up through the primary, the leading edge of the secondary becomes decelerated. This triggers fragmentation of the secondary disc to form two new condensations. These merge to form a binary companion to the secondary star. As this binary orbits, it sweeps out a gap in the surrounding material. In the meantime, the interaction has caused the primary disc to fragment to produce four new condensations. Figure \ref{fig:dd10b} shows the fragmentation of the primary disc viewed from above. The two frames correspond to the last two frames of Fig.~\ref{fig:dd10a}.

The secondary has a circumstellar disc of radius 30AU and mass 0.16$\mbox{M}_{\sun}$. Its binary companion is of mass 0.14$\mbox{M}_{\sun}$, and their orbit has eccentricity 0.16 and periastron 90AU. The primary has a tidally-truncated circumstellar disc of mass 0.15$\mbox{M}_{\sun}$ and radius 60AU, and is in a small-$\cal{N}$ cluster with companions of mass 0.02, 0.04, 0.06 and 0.07$\mbox{M}_{\sun}$. The cluster and the binary are bound to each other, with eccentricity 0.19 and periastron 930AU. The circumstellar disc around the primary is untilted by the encounter, but the outer disc and the orbits of the newly-formed companions have been twisted by approximately $12^{\circ}$.

\begin{figure*}
\plottwo{\figpath/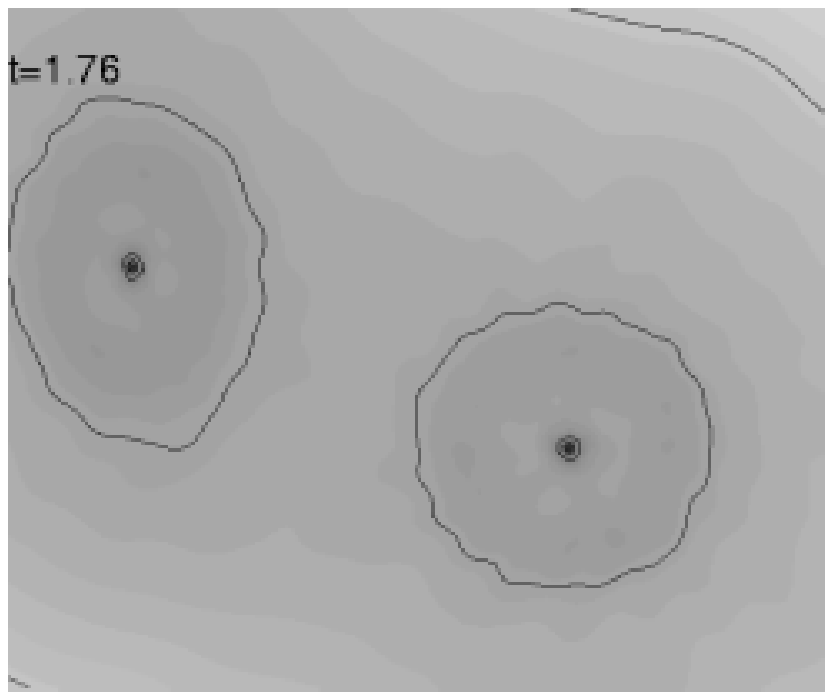}{\figpath/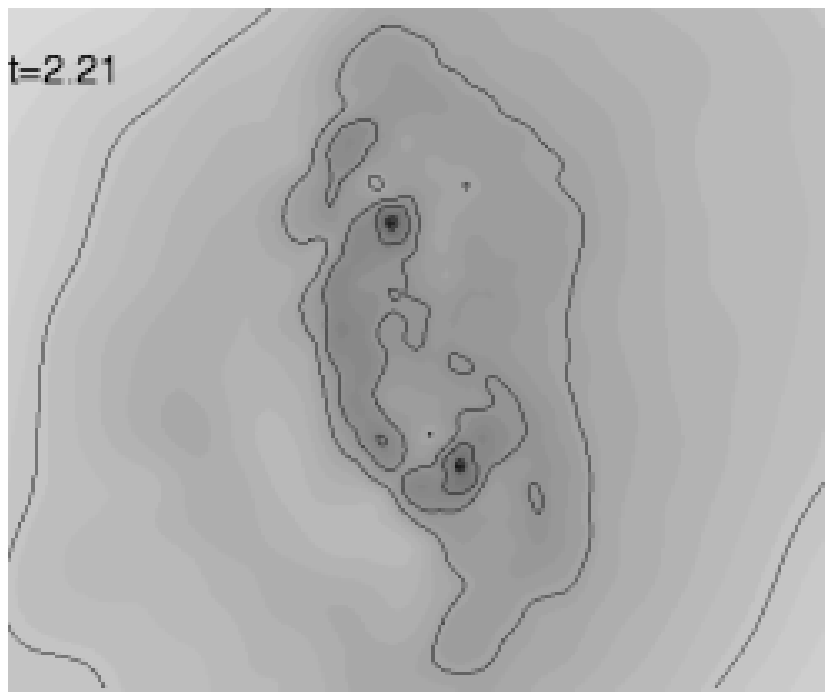}\\
\vspace{6pt}
\plottwo{\figpath/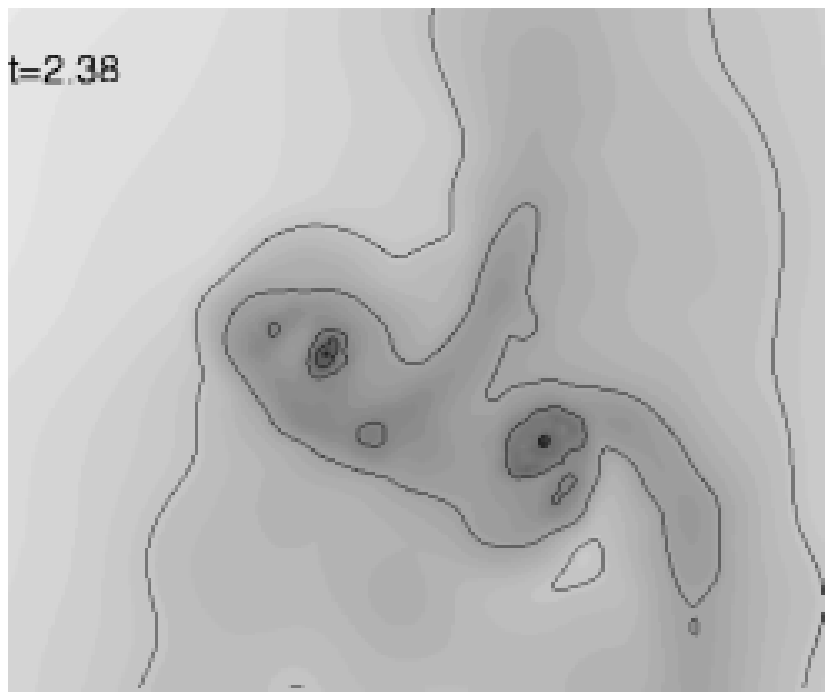}{\figpath/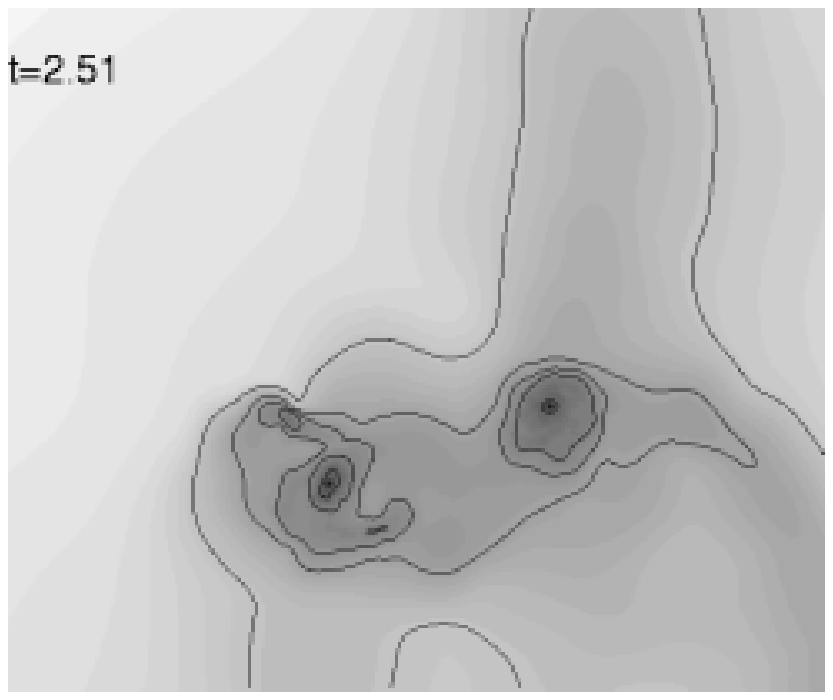}\\
\vspace{6pt}
\plottwo{\figpath/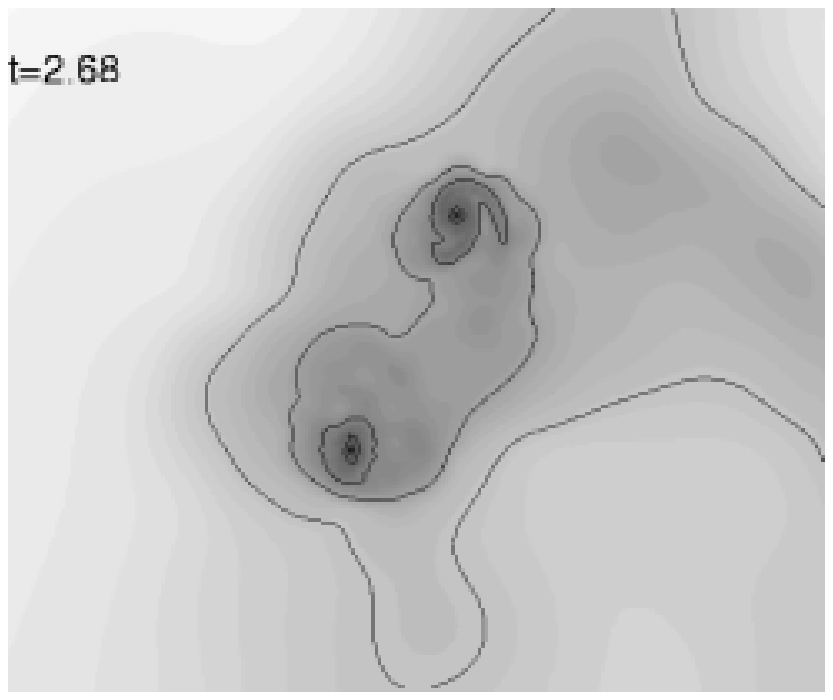}{\figpath/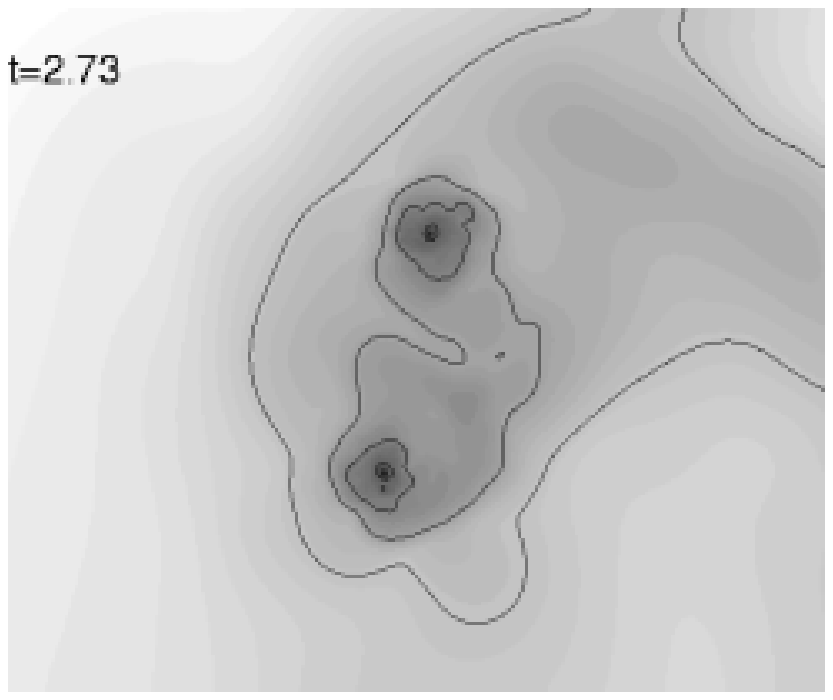}\\
\caption[Simulation dd13 : $\phi=3\pi/4$, $r_{peri}=500 \mbox{AU}$. 2000 $\times$ 1500 AU region.]{\label{fig:dd13}Simulation dd13 : $\phi=3\pi/4$, $r_{peri}=500 \mbox{AU}$. 2000 $\times$ 1500 AU region. Contour levels at [0.3, 3, 9 (frame 2, 4), 30, 300] $\mbox{g cm}^{-2}$. The primary is initially at the right-hand side of the figure, rotating clockwise and in the plane $z=0$. The secondary is initially below and to the left of the primary. Both the orbit and the secondary are inclined at $3\pi/4$ to the primary. Viewed from this angle, the orbit is in an anti-clockwise sense. The encounter does not lead to fragmentation of the discs, but the primary and secondary are captured into a binary system.}
\end{figure*}

\subsection{run dd13 : $\phi=3\pi/4$}

In runs dd13-16, the orbit and the secondary disc are inclined at 135 degrees to the primary disc. The results of run dd13 are shown in Fig.~\ref{fig:dd13}. The periastron is 500AU. The secondary is initially below the primary. As it penetrates through the primary disc, spiral arms form which join the secondary disc and the primary star, and likewise the primary disc and secondary star (frame 2). Due to the relative inclinations of the discs, these arms both lie on the same side. The outer regions of the discs then form into trailing arms, while the original arms dissipate as the stars move apart (frame 3). The central discs around each star undergo strong spiral-arm instabilities without fragmenting (frames 4 \& 5), before settling down to a steady state (frame 6). At the end of the encounter, the primary has a circumstellar disc of mass 0.44$\mbox{M}_{\sun}$ and radius 60AU, and the secondary has a disc of mass 0.37$\mbox{M}_{\sun}$ and radius 125AU. The two stars are in a binary orbit with eccentricity 0.32 and periastron 400AU.

\begin{figure*}
\plottwo{\figpath/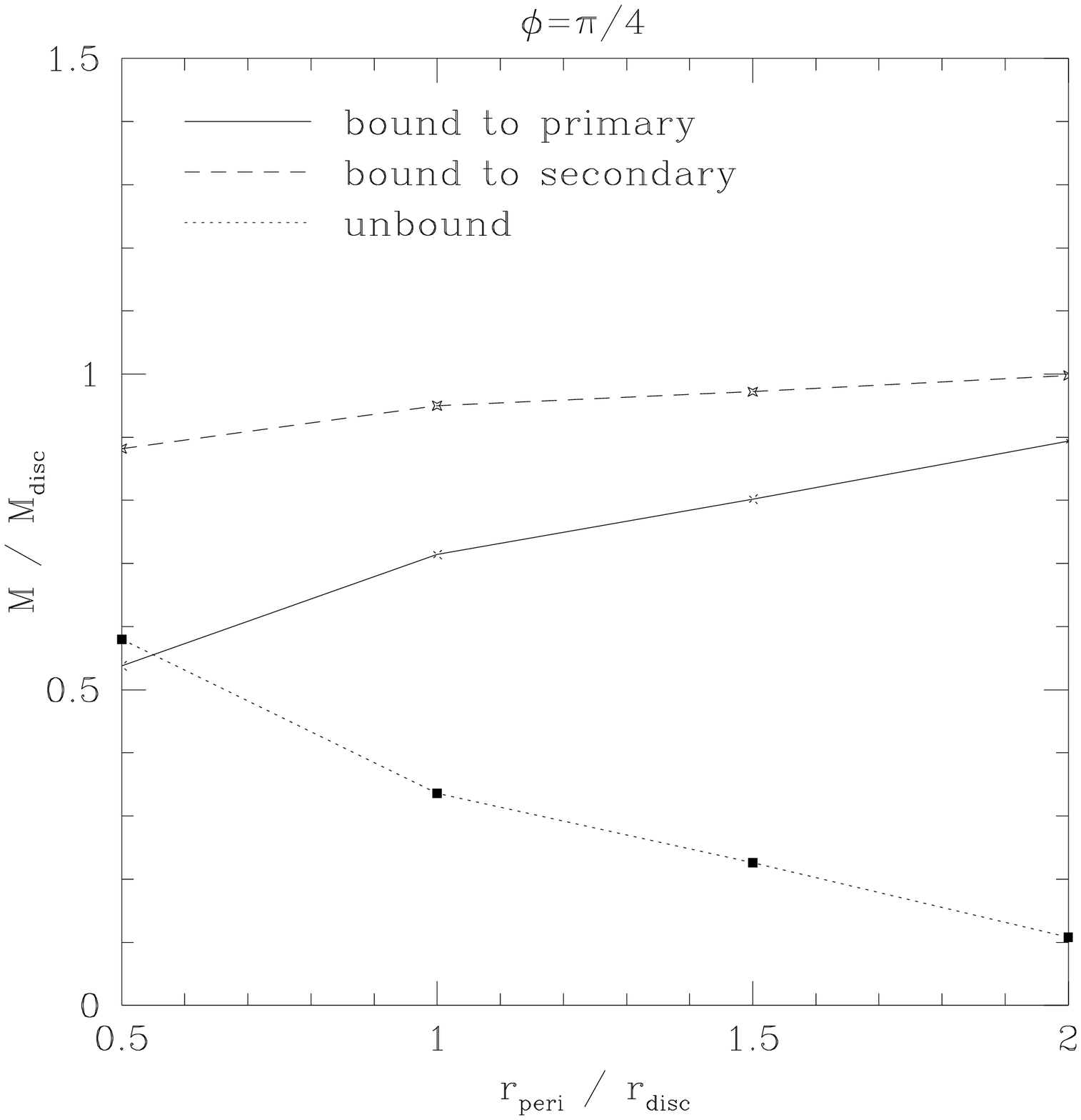}{\figpath/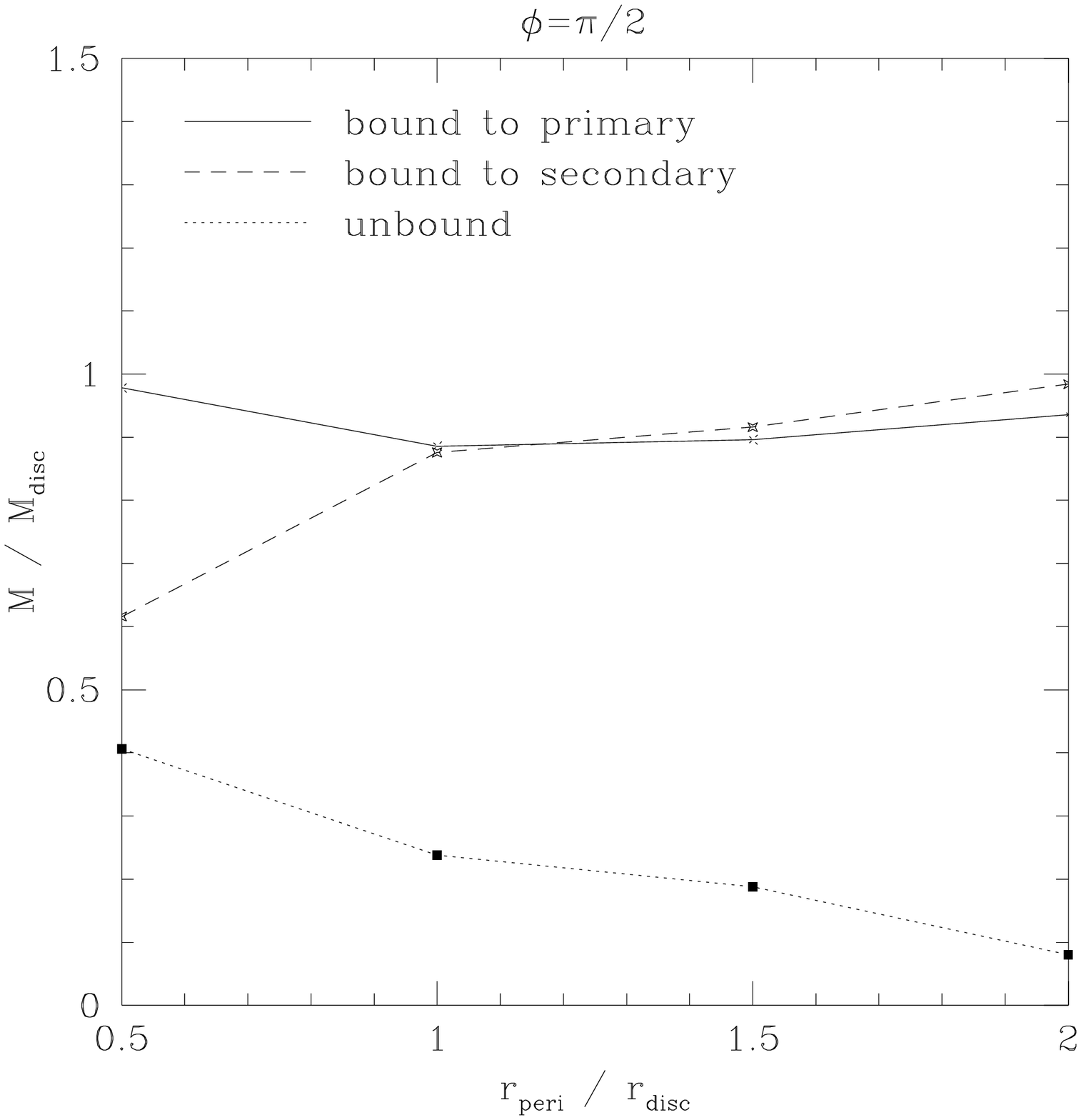}\\
\vspace{6pt}
\plottwo{\figpath/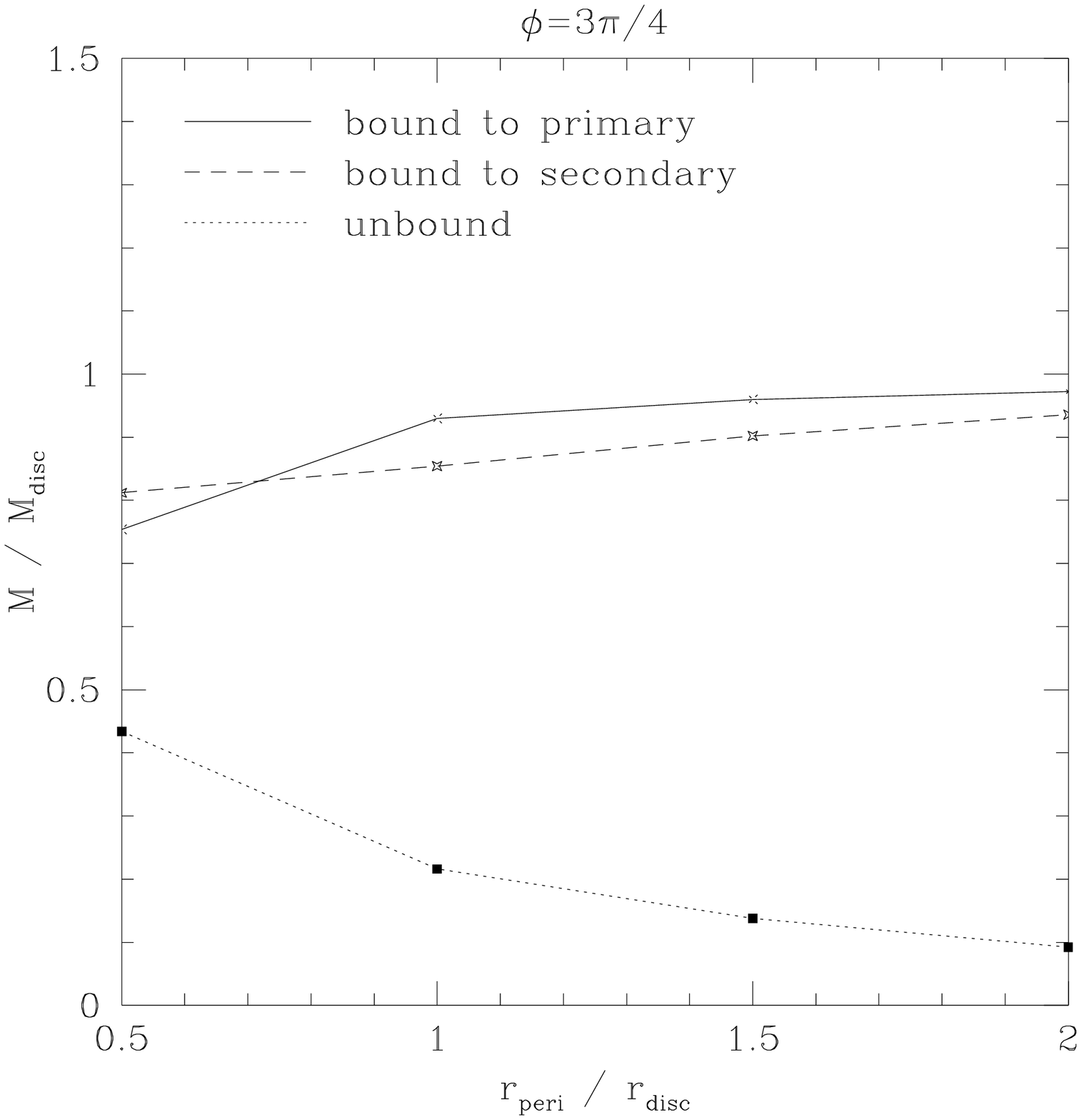}{\figpath/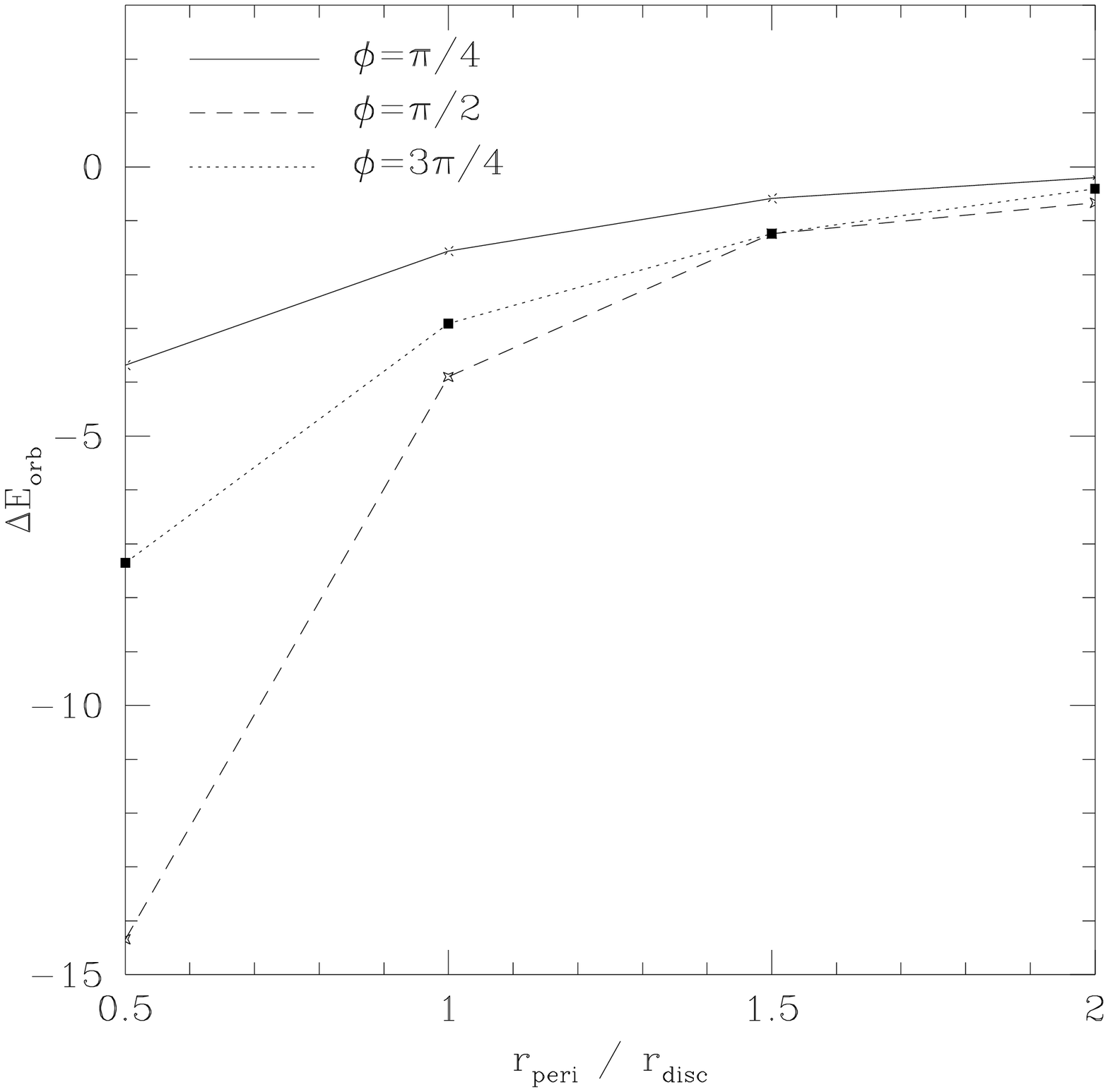}\\
\caption[Mass and energy transfer for non-coplanar disc-disc encounters]{\label{fig:ncpdmass}Frames 1-3 show the fraction of the disc mass that ends up bound to the primary star, bound to the secondary star and unbound, for $\phi=\pi/4$, $\pi/2$ and $3\pi/4$ encounters respectively. Frame four shows the change in energy of the original orbit for all non-coplanar encounters.}
\end{figure*}

\section{Results}

 In coplanar disc-disc encounters, disc material is swept up into a shock layer that then fragments to produce new protostars. The remaining discs may later undergo more fragmentation via spiral arm instabilities triggered by the encounter. In the case of non-coplanar encounters, shocks do not play a major part in the evolution of the system. Instead, the spiral-arm instabilities caused by the interaction are the sole source of fragmentation and energy transfer from the orbit of the discs.

Frames 1-3 of Fig.~\ref{fig:ncpdmass} show the mass that ends up bound to each of the original stars, and the mass that becomes unbound from the system, for each of the non-coplanar encounters. The mass that is bound to a star may be either in the form of a circumstellar disc, or bound up in newly-formed companions. The results for the corresponding disc-star encounters are given in Paper I. A similar amount of mass is unbound from the system for the star-disc and disc-disc encounters, but in the disc-disc encounters the secondary does not capture any material from the primary.

\subsection{Energy exchange}

Frame four of Fig.~\ref{fig:ncpdmass} shows the change in the energy of the orbit of the discs due to the encounter. The energy change is given in code units, in which the total binding energy of the disc is 6.6 and the binding energy outside $0.5r_{disc}$ is 0.82. As expected, the encounters with $\phi=3\pi/4$, which have a retrograde element to the orbit, remove more energy from the orbit than those with $\phi=\pi/4$. However, the most dissipative non-coplanar encounters, as well as being the ones that cause the most fragmentation, are the orthogonal encounters. This is in contrast to the corresponding disc-star encounters, in which the orthogonal encounters remove the least energy from the orbit.

\begin{figure*}
\plottwo{\figpath/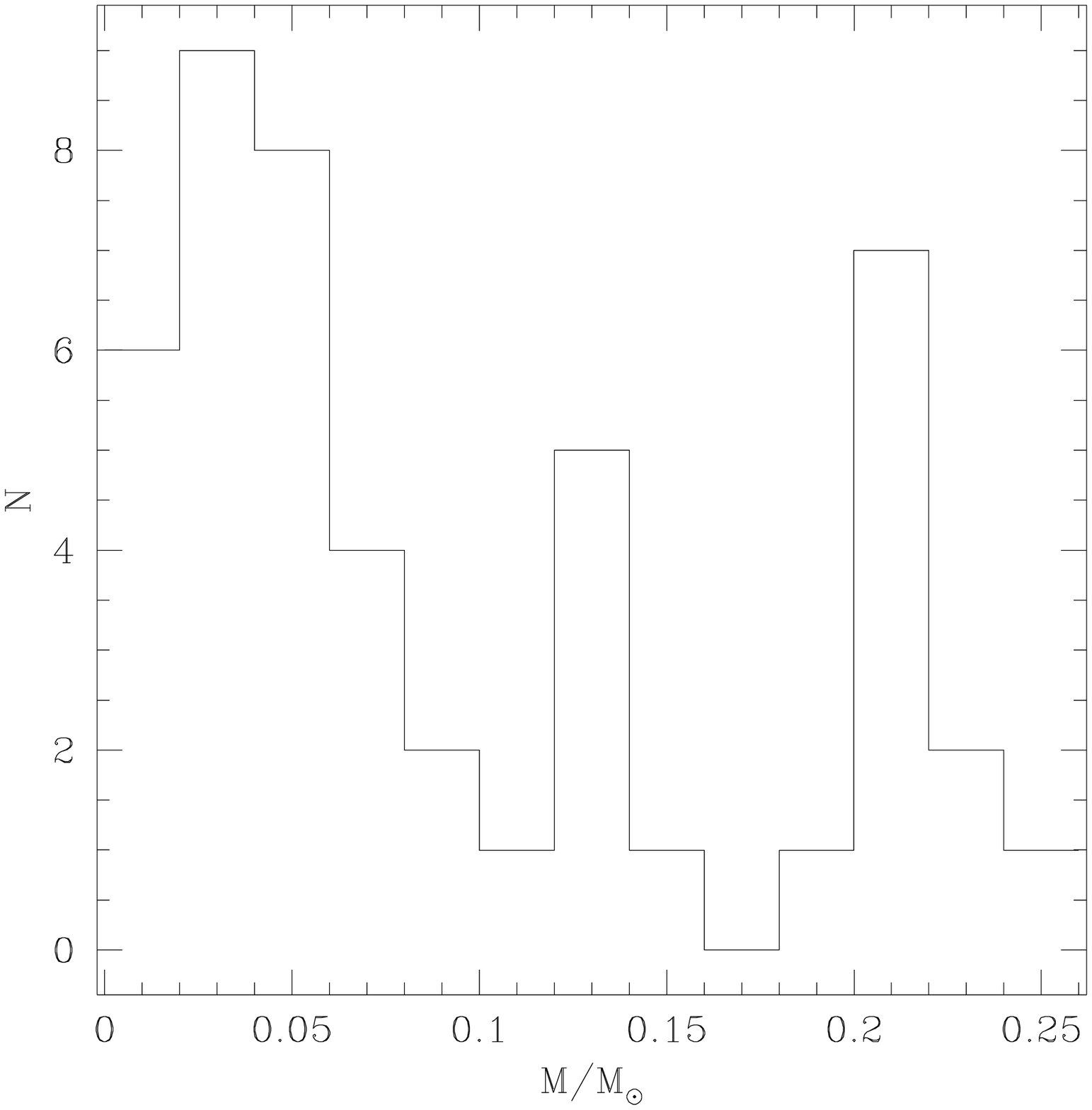}{\figpath/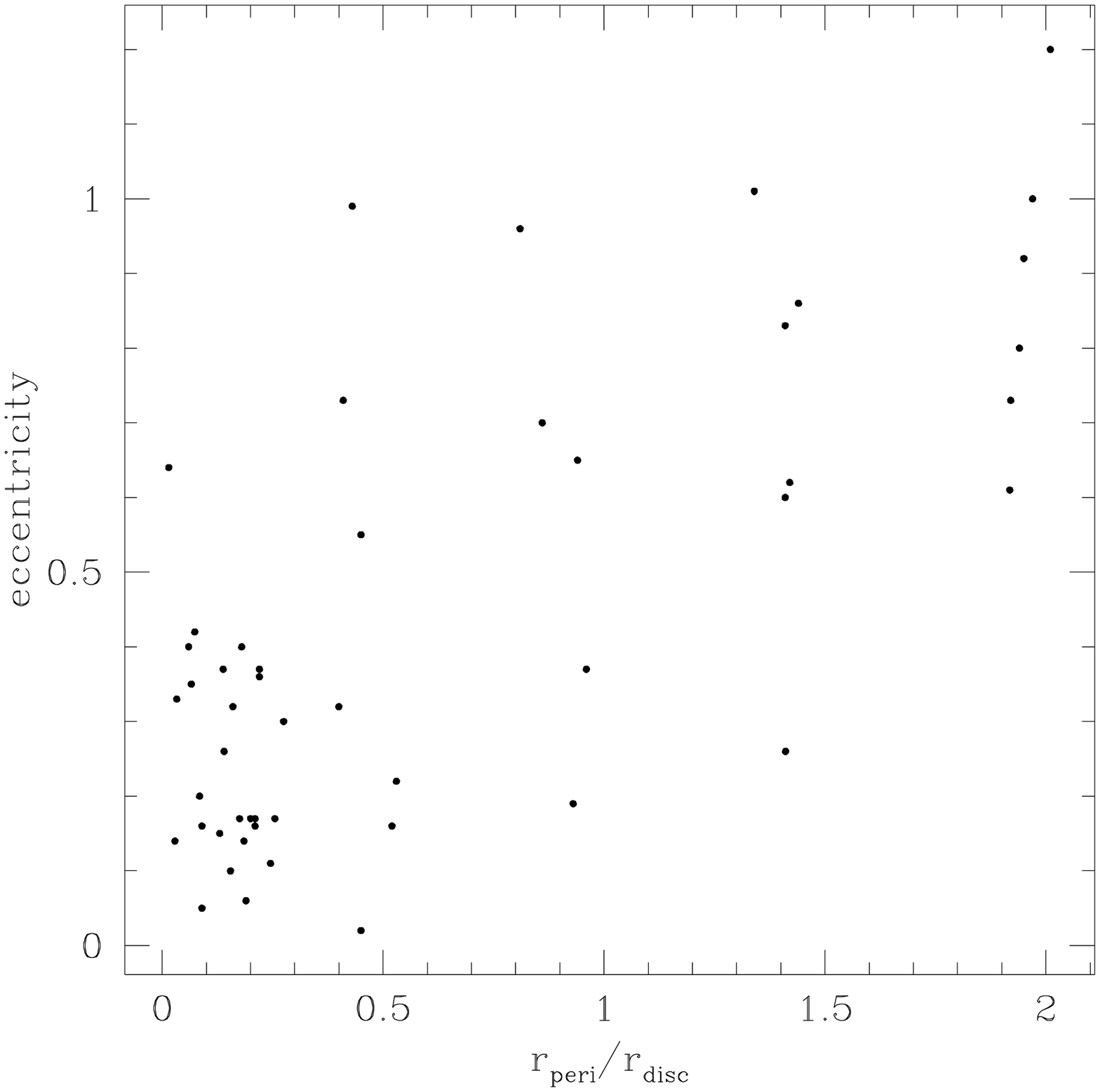}\\
\caption[Distribution of masses and orbital eccentricities]{\label{fig:orbits}Distribution of masses of newly-formed stars, and orbital eccentricities as a function of periastron after disc-disc encounters}
\end{figure*}

The energy removed from the orbit is much greater for disc-disc encounters than for the corresponding star-disc encounters, the results for which are shown in Fig.\ 6 of Paper I. For encounters with $\phi=\pi/4$, the disc-disc encounters remove 2-3 times as much energy from the orbit as do the corresponding disc-star encounters; for those with $\phi=\pi/2$ and $\phi=3\pi/4$, the ratio reaches as high as 20. Due to the alignment of the spin of the secondary disc with that of the orbit in all of the encounters presented here, energy will be transferred to the orbit via a corotation resonance. It is therefore likely that the results here provide a lower limit to the energy dissipated from the orbit during such encounters. Encounters with random inclinations, in which there is no resonance, will remove even more energy from the orbit of the stars. As a result of this, capture is likely to be a more efficient binary formation mechanism for disc-disc encounters than for star-disc encounters.

\subsection{Properties of new stars}

As well as removing energy from the orbit of the stars, the gravitational instabilities in the disc triggered by the encounter can lead to fragmentation of the disc and the formation of new companions to the original stars. For the non-coplanar star-disc encounters, fragmentation occurred in just one out of twelve simulations, leading to the formation of two companions to the primary star. For the corresponding disc-disc encounters, fragmentation occurs in eight of the 12 cases, producing a total of 16 new stars. Of those 16 stars, seven are produced by fragmentation of the primary disc, and nine by fragmentation of the secondary, so that it seems there is no particular preference for which disc fragments.

The left-hand frame of Fig.~\ref{fig:orbits} shows the final distribution of the masses of the protostars that are formed during the disc-disc encounters. Most of the fragments form with masses of around 0.04$\mbox{M}_{\sun}$. Condensations that form by fragmentation of the shock layer in coplanar encounters accrete from the discs and end up with masses of approximately 0.12$\mbox{M}_{\sun}$, whilst the condensations formed in the SOA\footnote{SOA stands for {\it spin orbit antiparallel}, i.e. the two spins are parallel to one another and antiparallel to the orbit (see Paper II).} encounters accrete the most material and end up with masses typically of 0.2$\mbox{M}_{\sun}$. In most cases there is still disc material that will be accreted, and in a physical situation there will also be a remnant infalling envelope, so that this distribution of masses may evolve greatly, and therefore may not be particularly close to the final outcome. The right-hand frame of Fig.~\ref{fig:orbits} shows the distribution of eccentricities as a function of periastron for all of the systems that are formed during disc-disc encounters. It can be seen that the encounters tend to change the eccentricities but not the periastra of the original orbits, leading to a clustering of points at $r_{peri}/r_{disc}$=0.5, 1.0, 1.5, 2.0. The orbits of the stars that are formed during the encounter have smaller eccentricities and periastra. Although there is a lot of scatter in the data, a general trend can be seen from having low eccentricities at low periastra to high eccentricities at larger periastra, as is observed amongst both pre-main-sequence and main sequence stars (Mathieu 1994\nocite{mathieu94}, Duquennoy \& Mayor 1991\nocite{duquennoy:mayor}). 

\begin{figure}
\plotsmall{\figpath/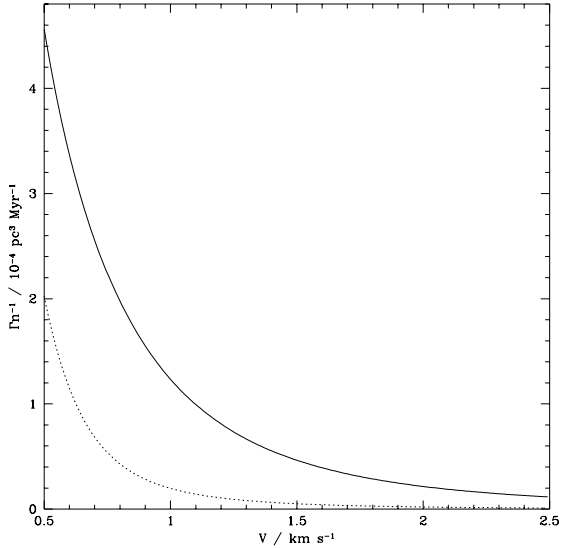}\\
\caption[Capture rate for disc-disc encounters]{\label{fig:disccap}Capture rate per unit number density for disc-disc encounters, as a function of velocity dispersion. The solid line shows the rate for randomly distributed inclinations, and the dots show the rate corrected for previous disruption of discs. If all encounters are SOP then the capture rate is negligible.}
\end{figure}

\begin{figure}
\plotsmall{\figpath/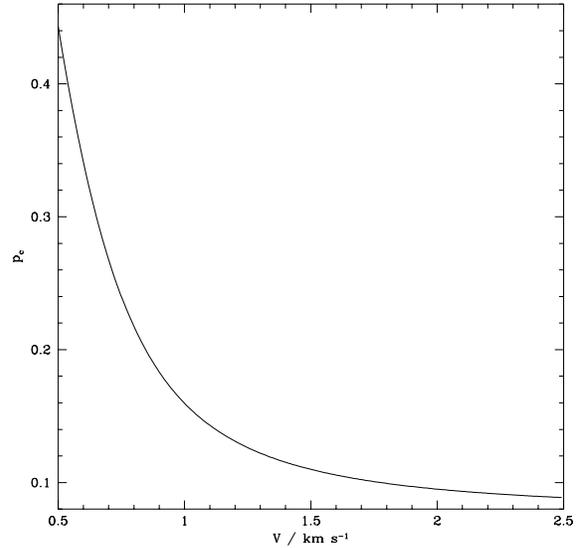}\\
\caption[Probability of capture for disc-disc encounter]{\label{fig:ddcap}Mean probability of capture during a disc-disc encounter for a randomly-inclined encounter, as a function of velocity dispersion.}
\end{figure}

\begin{table*}
\begin{center}
\begin{tabular}{|c|c|c|c|c|c|} \hline
Region  & $n / \mbox{pc}^{-3}$ & $V / \mbox{km s}^{-1}$ & $\Gamma_{c} / \mbox{Myr}^{-1}$ & $\Gamma_{c}^{'} / \mbox{Myr}^{-1}$ \\ \hline \hline
Trapezium (centre) & $10^{4}$ & 1.5 &  0.6 & 0.03\\ \hline
Trapezium & $2 \times 10^{3}$ & 1.5 &  0.12 & $6 \times 10^{-3}$\\ \hline
Open cluster & $10^{2}$ & 1 &  0.013 &  $2 \times 10^{-3}$ \\ \hline
GMC & 12 & 2 &  $3 \times 10^{-4}$ & $5 \times 10^{-6}$ \\ \hline
\end{tabular}
\end{center}
\caption[Capture rates for disc-disc encounters]
{\label{table:ddcap}Capture rates for disc-disc encounters in different star-forming environments}
\end{table*}

\begin{table}
\begin{center}
\begin{tabular}{|c||c|c|c|c|c|c|} \hline
$r_{peri}$  & {  SOP   }& $\phi=\frac{\pi}{4}$ & $\phi=\frac{\pi}{2}$ & $\phi=\frac{3\pi}{4}$ & SOM & SOA \\ \hline \hline
500 & 2 & 0 & 1 & 0 & 3 & 0 \\ \hline
1000 & 3 & 3 &  5 & 1 & 7 & 2 \\ \hline
1500 & 3 & 0 &  2 & 1 & 3 & 2 \\ \hline
2000 & 2 & 1 &  2 & 0 & 1 & 2 \\ \hline
\end{tabular}
\end{center}
\caption[Companion formation rates for disc-disc encounters]
{\label{table:ddcomp}Number of companions formed by disc fragmentation}
\end{table}

\subsection{Capture rates}

In order to attempt to ascertain how important capture and fragmentation are for the formation of multiple systems, the calculations of Paper I are repeated for the disc-disc encounters. We calculate both the capture rate and the rate of formation of companions by fragmentation. As fragmentation occurs for almost all of the disc-disc encounters, the rate of capture events will not in general be the rate of formation of binaries by this mechanism, but rather the rate of capture of single stars and multiple systems into hierarchical configurations. We consider only encounters in which the spin of the secondary disc is parallel to that of the orbit, so that for a coplanar encounter with inclination $\phi=\pi$, we use the SOM\footnote{SOM stands for {\it spin orbit mixed}, i.e. one disc spins parallel to the orbit, and the other spins antiparallel to the orbit (see Paper II).} rather than the SOA results.

Figure \ref{fig:disccap} shows the capture rate per unit number density for disc-disc interactions, $\Gamma_{c}$, as a function of velocity dispersion. It also shows the rate corrected for previous encounters, $\Gamma_{c}^{'}$ which assumes that an encounter with $r_{peri}<2r_{disc}$ completely destroys both discs. This information is then used to calculate capture rates in the star-forming environments considered in Clarke \& Pringle \shortcite{clarke:pringle91a} and in Paper I. The rates are given in Table \ref{table:ddcap}. It can be seen that in typical star-forming environments such as an open cluster or a GMC, capture does not happen at a significant rate, mainly due to the low number of encounters that occur. The rates corrected for previous encounters are also much lower than the uncorrected rates, due to the fact that disc-disc encounters destroy the discs out to a greater periastron than disc-star encounters.

\begin{figure}
\plotsmall{\figpath/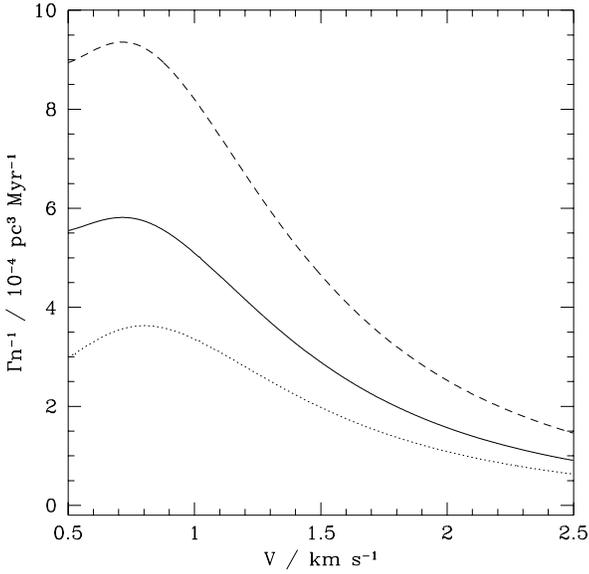}\\
\caption[Companion formation rate for disc-disc encounters]{\label{fig:disccomp}Rate of formation of companion stars, per unit number density for disc-disc encounters, as a function of velocity dispersion. The solid line shows the rate for randomly distributed inclinations, the long dashes show the rate assuming all encounters are prograde coplanar, and the short dashes show the rate corrected for previous disruption of discs.}
\end{figure}

\begin{table*}
\begin{center}
\begin{tabular}{|c|c|c|c|c|c|} \hline
Region  & $n / \mbox{pc}^{-3}$ & $V / \mbox{km s}^{-1}$ & $\Gamma_{f} / \mbox{Myr}^{-1}$ & $\Gamma_{f}^{'} / \mbox{Myr}^{-1}$ & $\Gamma_{p} / \mbox{Myr}^{-1}$ \\ \hline \hline
Trapezium (centre) & $10^{4}$ & 1.5 & 3.0 & 2.0 & 4.8 \\ \hline
Trapezium & $2 \times 10^{3}$ & 1.5 & 0.6 & 0.4 & 1.0 \\ \hline
Open cluster & $10^{2}$ & 1 & 0.05 & 0.03 & 0.08 \\ \hline
GMC & 12 & 2 & $2 \times 10^{-3}$ & $1 \times 10^{-3}$ & $3 \times 10^{-3}$ \\ \hline
\end{tabular}
\end{center}
\caption[Companion formation rates for disc-disc encounters]
{\label{table:ddbin}Rate of formation of companions for disc-disc encounters in different star-forming environments. $\Gamma_{f}$ is the uncorrected rate assuming random inclinations for encounters, $\Gamma_{f}^{'}$ is the corresponding corrected rate, and $\Gamma_{p}$ is the uncorrected rate if all encounters are coplanar prograde.}
\end{table*}

If star formation is dynamically-triggered, however, for instance by the collision between two clumps within a molecular cloud, then it is expected that almost every protostar formed will undergo at least one interaction \cite{turner}. The probability of capture during an encounter is then $p_{c}=\Gamma_{c}/\Gamma_{h}$, where $\Gamma_{h}$ is the rate of encounters with $r_{peri}<2 r_{disc}$. Figure \ref{fig:ddcap} shows $p_{c}$ as a function of the velocity dispersion. In approximately 10\% of cases, the stars will be bound to each other after the encounter. Given that fragmentation occurs in most of the disc-disc encounters, it will usually not be two individual stars that become bound to each other in a binary following capture. Rather, at least one of the two stars will already have one or more companions, so that the outcome will be a hierarchical system.

If all of the interactions are SOP\footnote{SOP stands for {\it spin orbit parallel}, i.e. both discs spin parallel to the orbit (see Paper II).} encounters, then the capture rate is negligible, due to the resonances that tend to unbind the stars from each other.

\subsection{Fragmentation rates}

The rate of formation of companions has also been calculated using the same method as in Paper I. Table \ref{table:ddcomp} lists the number of companions formed during each of the disc-disc encounters simulated. When calculating the rate of formation of companions, we divide these numbers by two to obtain the number of companions formed per disc in the simulations. Figure \ref{fig:disccomp} shows the rate of formation of companions per unit number density. This information is then used to calculate the companion formation rates in typical star-forming regions, given in Table \ref{table:ddbin}. In high-density regions like the Trapezium, fragmentation during disc-disc collisions can produce a high number of companions, so that for instance in the very core of the Trapezium, a star will have on average 2 to 5 companions after 1Myr. In lower density regions, the rates are much lower due to the low number of interactions.

\begin{figure}
\plotsmall{\figpath/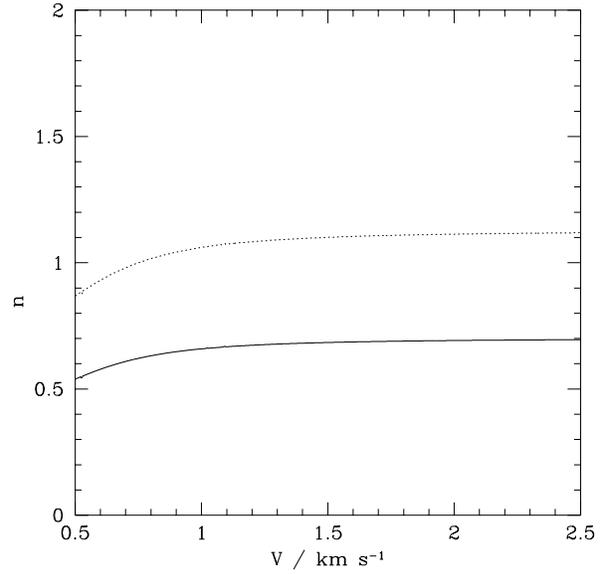}\\
\caption[Mean companion formation rate for a disc-disc encounter]{\label{fig:ddnfrag}Mean number of companions formed in a disc-disc encounter, as a function of velocity dispersion. The solid line shows the number for randomly distributed inclinations, the dots show the number assuming all encounters are prograde and coplanar.}
\end{figure}

\section{Discussion}

\subsection{Disc-star interactions}

Disc-star interactions provide a mechanism for removing energy from the orbit of the two interacting stars, for truncating the disc, and for triggering the fragmentation of the disc to produce new stars. The results presented in Paper I are in general agreement with those of previous workers, with prograde coplanar interactions being the most destructive to the disc, but retrograde coplanar interactions removing the most energy from the orbit. The evolution of the penetrating, retrograde, coplanar interactions is dominated by the formation of a trailing shock behind the perturbing star that dissipates large amounts of energy and leads to the fragmentation of the disc. For other penetrating interactions, the energy removed from the orbit is approximately twice the binding energy of the disc outside of periastron.

If star formation within typical star-forming regions is distributed, then the capture of the two stars into a binary system due to the energy dissipated in an interaction happens at a negligible rate, primarily due to the low number of discs that experience interactions. Even if every disc undergoes an interaction, as might happen during dynamically-triggered star formation, then the capture rate is extremely low unless the stars initially have a very low velocity dispersion. If the majority of interactions are prograde, coplanar encounters, as predicted for stars that are formed from the same core or disc, or that are formed during the collision of two clumps within a molecular cloud, then the interactions will tend to unbind the stars, and no binaries will form by capture.

Disc-star interactions, as well as removing energy from the orbit of the interacting stars, can also trigger the fragmentation of the disc via gravitational spiral-arm instabilities, leading to the formation of one or two new stars that are bound to the original star. If every disc undergoes one disc-star interaction, then approximately 30\% of stars will end up in multiple systems, rising to about 60\% if all of the interactions are prograde and coplanar. Thus disc-star interactions can give rise to a large number of multiple systems, formed primarily by fragmentation of the disc triggered by the interaction, rather than by capture.

\subsection{Disc-disc interactions}

If the star formation within a region is coeval, as will be the case if it is dynamically triggered, then most encounters are likely to be disc-disc interactions, rather than star-disc interactions. For disc-disc interactions, the evolution of the system is almost always dominated by the fragmentation of the disc to produce multiple new companions to the original stars. This fragmentation occurs in two forms. The first, which happens for non-coplanar encounters, is via the same gravitational spiral-arm instabilities as for the star-disc encounters. In the second mechanism, which dominates for the coplanar disc-disc encounters, the disc material between the two interacting stars is swept up into a shock layer that fragments to produce multiple new stars. Most often these form binary or triple systems with one of the original stars, but occasionally larger clusters can form. 

Disc-disc encounters form 2-3 times as many companions as star-disc encounters. Capture will occur in approximately 15\% of disc-disc encounters, a significantly higher rate than for star-disc encounters. Given the high frequency of fragmentation, in most cases capture will produce a small-${\cal N}$ cluster of several stars, and so the process of disc-disc interaction can repeat itself hierarchically.

\subsection{Disc masses}

The calculations presented in this paper and its two companions have demonstrated that under suitable circumstances (i) interactions between massive protostellar discs, and/or between a massive protostellar disc and a naked star, can spawn additional protostars from the disc material; and (ii) most of the resulting protostars are in binary or higher multiple systems, with orbits smaller than the periastron of the original interacting pair. Therefore, if these processes can, starting with a relatively wide interacting pair, repeat themselves in an hierarchical cascade, it may be possible to create ever closer systems, and hence to populate the entire observed range of binary separations. However, this can only happen if the protostellar discs involved remain massive, so that there is a sufficient reservoir of material from which to form new protostars.

This requirement places at least two conditions on the evolution of protostellar discs. The first is that in an isolated protostellar disc the rate at which matter looses angular momentum and spirals onto the central star must be slow. Specifically, the time-scale $t_{\mbox{\tiny visc}}$ on which an isolated protostellar disc accretes onto its central star should be longer than the mean time $t_{\mbox{\tiny int}}$ between interactions. There seems no reason to suppose that this condition is not met routinely in star-formation regions. $t_{\mbox{\tiny visc}}$ is essentially the time-scale on which angular momentum is redistributed in the disc, due to turbulent viscosity, or magnetic stresses, or global gravitational instabilities; $t_{\mbox{\tiny visc}}$ is likely to be significantly longer than the spin period of the disc $t_{\mbox{\tiny spin}}$. By contrast, when clusters of protostellar discs form (by fragmentation) from the same collapsing core, $t_{\mbox{\tiny int}}$ should be comparable with $t_{\mbox{\tiny spin}}$. Indeed, it is possible that most accretion onto the central star occurs as a result of non-linear instabilities triggered by interaction with another protostar; and that between interactions an isolated protostellar disc evolves secularly, with matter accreting only very slowly onto the central star, and the remaining disc becoming increasingly extended -- and therefore increasingly prone to interaction -- until it is involved in (another) impulsive interaction.

The second condition is that protostellar discs must be replenished by infall from the surrounding gas. Otherwise the hierarchical cascade will tend to deliver protostars with ever-decreasing, and eventually sub-stellar, masses. Again this condition is likely to be fulfilled. Both theory and observation seem to be agreed that the gravitational instability which creates the original protostellar discs is likely to be followed by continuing infall. As long as most of the infalling material has sufficient angular momentum relative to the individual central stars, it will accrete onto their discs, thereby replenishing them in preparation for the next interaction.

Thus, as long as (i) the evolution of an isolated protostellar disc is sufficiently slow, and (ii) protostellar discs are replenished by infall, the processes which we have modelled here could operate frequently, and in an hierarchical cascade to generate a wide range of binary (and higher multiple) systems.

\section{Conclusions}

If star formation is not quiescent, but is dynamically triggered, then it is likely that almost every protostar undergoes at least one interaction whilst a large fraction of its mass is still in an extended disc. Figure \ref{fig:ddnfrag} shows the mean number of companions formed in a disc-disc interaction. If every protostellar disc undergoes an interaction which has a random inclination, then after the encounter the star will on average have 0.6 companions. If all of the interactions are coplanar SOP, then the mean number of companions is approximately double that for randomly-inclined interactions. Some of the stars will be in binary systems, some in higher multiples, and some will stay as single stars. Many of the higher multiple systems will be in unstable configurations, and some stars will undergo slingshot events and be ejected from the birth site.

For non-coplanar disc-disc interactions, the evolution of the system is dominated by spiral-arm gravitational instabilities. These instabilities can lead to the fragmentation of the discs, and the formation of new companions to the original stars. This occurs in two-thirds of the non-coplanar interactions modelled here.

We have calculated the rates of capture due to disc-disc interactions. Given the large fraction of cases in which fragmentation occurs, any capture is likely to be of a multiple system into a hierarchical grouping. We find that for randomly-oriented interactions, capture will occur in about 20\% of cases; if, however, all of the interactions are coplanar and spin-orbit parallel, then the rate of capture is negligible.

We believe that the majority of protostellar discs will undergo at least one disc-disc interaction during their lifetime. The results presented here indicate that such interactions lead to the formation of new protostellar discs in closer multiple systems. Therefore the  compound effect could be to propagate the star-formation process from a few initial massive protostellar discs in wide binary systems, to lower masses and smaller separations, thereby populating most of the observed range of stellar masses and binary separations.

\section*{Acknowledgements}

SJW and ASB acknowledge the receipt of University of Wales studentships. NF acknowledges the receipt of a PPARC studentship. HB and SJW acknowledge the support of a PPARC post-doctoral grant GR/K94157. This work was supported by grants GR/K94140 and GR/L29996 from PPARC.

\end{document}